\documentclass[showpacs,preprintnumbers,amsmath,amssymb,12pt,floatfix,epsfig]{revtex4}
\usepackage{graphicx}
\begin{document}
\draft
\title{Strange quark contributions to neutrino and antineutrino nucleus scattering
via neutral current in quasi-elastic region }
\author{K. S. Kim$^{1)}$, Myung-Ki Cheoun$^{2)}$
\footnote{Corresponding author : cheoun@ssu.ac.kr}}
\address{1)School of Liberal Arts and Science, Korea Aerospace
University, Koyang 200-1, Korea \\ 2)Department of Physics,
Soongsil University, Seoul, 156-743, Korea}

\begin{abstract}
Strange quark contributions to the neutral current reaction in
neutrino(antineutrino) scattering are investigated on the nucleon
level and extended to the $^{12}$C target nucleus through the
neutrino-induced knocked-out nucleon process in the quasi-elastic
region within the framework of a relativistic single particle
model. The incident energy range between 500 MeV and 1.0 GeV is
used for the scattering. Effects of final state interactions for
the knocked-out nucleon are included by a relativistic optical
potential. We found that there exist some cancellation mechanisms
between strange quark contributions by protons and neutrons inside
nuclei in inclusive reactions, $A ( \nu ({\bar \nu}), \nu^{'}
({\bar \nu}^{'}) )$. As a result, the sensitivity of the strange
quark contents could be more salient on the asymmetry between
neutrino and antineutrino scattering in knocked-out nucleon
processes, $A ( \nu ({\bar \nu}), \nu^{'} ({\bar \nu}^{'}) N )$.
\end{abstract}
\narrowtext
\maketitle

Neutrino-nucleus $ (\nu - A)$  scattering has become to be widely
interested in different fields of physics such as astrophysics,
cosmology, particle, and nuclear physics. In particular, the
scattering of neutrinos$({\nu})$ and antineutrinos$({\bar \nu})$
on nuclei enabled us to obtain some invaluable clues on the
strangeness contents of the nucleon. Along this line, Brookhaven
National Laboratory (BNL) \cite{brook} reported that the value of
a strange axial vector form factor of the nucleon does not have
zero.

At intermediate $\nu ( {\bar \nu)}$ energies, there are many
theoretical works
\cite{garvey,hung,umino,alberico,jacowicz,giusti1,udias,madrid}
for the $ \nu - A $ scattering, for example, relativistic Fermi
gas (RFG) models \cite{hung,umino}, the relativistic plane wave
impulse approximation (RPWIA) \cite{alberico}, and
non-relativistic nuclear shell models \cite{jacowicz}.

One of recent developments is to extract model-independent and
target-independent predictions for the $ \nu - A $ scattering. For
instance, ref. \cite{barbaro,anto} exploited the superscaling
analysis (SuSA), which has been used for scaling analysis of
inclusive electron scattering. They showed that the SuSA method
yields predictions for neutrino scattering with the uncertainty
about 15 $\sim$ 20 \% level compared with RFG models. Moreover it
has been carried out not only in the quasi-elastic (QE) region but
also in the $\Delta$ region.

Apart from the nuclear models, the final state interaction (FSI)
of the knocked-out nucleon is one of important ingredients in
relevant reactions. Usually two different methods, the complex
optical potential \cite{giusti1} and the relativistic multiple
scattering by the Glauber approximation \cite{ghent}, are
exploited for the FSI. In particular, ref. \cite{nieves} shows
that the ratio of proton and neutron cross section in the QE
region is sensitive to the strange quark axial form factor of the
nucleon by using the Monte Carlo treatment of the rescattering.

In this paper, we investigate the strange quark contents on the
nucleon by considering the neutral current (NC) scattering on the
nucleon, and applying its results to the $ \nu - A$ scattering in
the QE region, where inelastic processes like pion production and
$\Delta$ resonance are excluded. Beyond the QE region, of course,
one has to include such inelastic processes. For example, ref.
\cite{leitner} showed that the contribution of $\Delta$ excitation
is comparable to that of the QE scattering at the neutrino
energies above 1 GeV.

Relativistic bound state wave functions for the nucleon are
obtained from solving a Dirac equation in the presence of strong
scalar and vector potentials based on the $\sigma - \omega$ model
\cite{horo}. In order to include the FSI we take a relativistic
optical potential \cite{clark}. Our nuclear model used here has
been successfully applied to the $A (e, e^{'})$ and $A (e, e^{'} N
)$ reactions \cite{kim1,kim2}. Incident neutrino (antineutrino)
energies are concerned in intermediate ranges between 500 MeV and
1.0 GeV.

We start from a weak current on the nucleon level, $J^{\mu}$,
which represents the Fourier transform of the nucleon current
density written as
\begin{equation}
J^{\mu}=\int {\bar \psi}_p {\hat {\bf J}}^{\mu} \psi_b e^{i{\bf
q}{\cdot}{\bf r}}d^3r,
\end{equation}
where ${\hat {\bf J}}^{\mu}$ is a free weak nucleon current
operator, and $\psi_{p}$ and $\psi_{b}$ are wave functions of the
knocked-out and the bound state nucleons, respectively. For a free
nucleon, the NC operator comprises the weak vector and the axial
vector form factors
\begin{equation}
{\hat {\bf J}}^{\mu}=F_{1}^V (Q^2){\gamma}^{\mu}+ F_{2}^V
(Q^2){\frac {i} {2M_N}}{\sigma}^{\mu\nu}q_{\nu} + G_A(Q^2)
\gamma^{\mu} \gamma^5 + {\frac {1} {2M_N}}G_P(Q^2) q^{\mu}
\gamma^5.
\end{equation}
By the conservation of the vector current (CVC) hypothesis with
the inclusion of an isoscalar strange quark contribution, $F_i^s$,
the vector form factors for protons and neutrons, $F_{i}^{V,~p(n)}
(Q^2)$, are expressed as \cite{giusti1}
\begin{eqnarray}
F_i^{V,~ p(n)} ( Q^2) &=&({\frac 1 2} - 2 \sin^2 \theta_W )
F_i^{p(n)} ( Q^2) - {\frac 1 2} F_i^{n(p)}( Q^2) -{\frac 1 2}
F_i^s ( Q^2)~ ,
\end{eqnarray}
where $\theta_W$ is the Weinberg angle given by $\sin^2 \theta_W =
0.2224$ \cite{udias}. Strange vector form factors are given as
\cite{garvey}
\begin{equation}
F_1^s(Q^2) = {\frac {F_1^s Q^2} {(1+\tau)(1+Q^2/M_V^2)^2}}~,
\;\;\;\;\; F_2^s(Q^2) = {\frac {F_2^s(0)}
{(1+\tau)(1+Q^2/M_V^2)^2}}~,
\end{equation}
where $\tau=Q^2/(4M_N^2)$, $M_V=0.843$ GeV,
$F_1^s=-<r_s^2>/6=0.53$ GeV$^{-2}$, and $F_2^s(0)=\mu_s$ is a
strange magnetic moment given by $\mu_s=-0.4$. Here we exclude
strange quark contributions to the electric form factor since
their effects turned out to be very small \cite{giusti1}. The
axial form factor is given by \cite{musolf}
\begin{eqnarray}
G_A (Q^2) &=&{\frac 1 2} (\mp g_A + g_A^s)/(1+Q^2/M_A^2)^2,
\label{gs}
\end{eqnarray}
where $g_A=1.262$, $M_A=1.032$ GeV. The $g_A^s=-0.19$ represents
the strange quark contents in the nucleon \cite{bernard}. $-(+)$
coming from the isospin dependence denotes the knocked-out proton
(neutron), respectively. Actually values of $F_1^s$ and $\mu_s$
have some ambiguity due to uncertainties persisting in the strange
form factors \cite{Budd03,Step06}. Physical parameters to the
strangeness are used from ref.\cite{giusti1} whose values are
constrained to pertinent experimental data, while other constants
are from ref.\cite{udias}. Since we take $+$ sign for $G_A (Q^2) $
in eq.(2), the axial form factor in eq.(5) is just negative to the
form factor elsewhere, for example, in ref.\cite{giusti1}

The induced pseudoscalar form factor is parameterized by the
Goldberger-Treimann relation
\begin{equation}
G_P(Q^2) = {\frac {2M_N} {Q^2+m^2_{\pi}}} G_A(Q^2),
\end{equation}
where $m_{\pi}$ is the pion mass. But the contribution of the
pseudoscalar form factor vanishes for the NC reaction because of
the negligible final lepton mass participating in this reaction.

The strange quark contributions are explicitly contained in
$F_i^s$ and $g_A^s$. Since the first term in Eq. (3) rarely
contributes due to the given Weinberg angle, the elastic cross
section on the proton, $\sigma ( \nu p \rightarrow \nu p)$, is
sensitive mainly on the $F_i^s$ and $g_A$ values. But measuring of
the cross section itself is not so easy experimentally task, so
that one usually resorts to the cross section ratio between the
proton and the neutron, $R_{p/n} = { {\sigma ( \nu p \rightarrow
\nu p ) } / { \sigma (\nu n \rightarrow \nu n )}}$. Measuring of
this ratio has also some difficulties in the neutron detection.
Therefore, the ratio $R_{NC/CC} = { {\sigma ( \nu p \rightarrow
\nu p ) } / { \sigma (\nu n \rightarrow \mu^- p )}}$ is suggested
as a plausible signal for the nucleon strangeness because the
denominator, the charged current (CC) cross section, is relatively
insensitive to the strangeness \cite{Vent04}.

In this paper, we investigate an alternative way to search for the
strangeness by using only NC reactions. It is an asymmetry on the
neutrino and antineutrino cross section, $A_{NC} = {
(\sigma_{NC}^{\nu} - \sigma_{NC}^{{\bar \nu}} ) /
(\sigma_{NC}^{\nu} + \sigma_{NC}^{{\bar \nu}}  )}$, where
$\sigma_{NC}^{\nu ({\bar \nu})}$ means differential cross sections
by $\nu$ and ${\bar \nu}$. If we use Sachs form factors usually
related by the nucleon form factors,
\begin{equation}
G_E (Q^2) = F_1 (Q^2) - { {Q^2} \over { 4 M^2}} F_2 (Q^2)~,G_M
(Q^2) = F_1 (Q^2) + F_2 (Q^2)~,
\end{equation}
the NC elastic neutrino(antineutrino) cross section on the nucleon
level is expressed in terms of the Sachs form factors
\cite{Albe02},
\begin{eqnarray}
{( {  {d \sigma  } \over {d Q^2  }} )}_{\nu ({\bar \nu}   )
}^{NC}&=& { {G_F^2 } \over {2 \pi }} [ { 1 \over 2} y^2 { (G_M
)}^2 + ( 1 - y - { { M }\over {2 E_{\nu} }} y ) { {{ (G_E
)}^2 + { E_{\nu} \over 2 M} y { (G_M )}^2 } \over {1 + {  E_{\nu} \over 2 M} y }} \\
\nonumber & & + ( { 1 \over 2} y^2 + 1 - y + { { M } \over {2
E_{\nu} }} y ){ (G_A )}^2 \mp 2 y ( 1 - { 1 \over 2 } y ) G_M G_A
]~.
\end{eqnarray}
Here $E_{\nu}$ is the energy of incident $\nu ({\bar \nu})$ in the
laboratory frame, and $y = { { p \cdot q } / { p \cdot k }} = { {
Q^2 } / {2 p \cdot k }}$ with $k,p$ and $q$, initial 4 momenta of
$\nu ({\bar \nu})$ and target nucleon, and 4 momentum transfer to
the nucleon, respectively. $\mp$ corresponds to the cases of the
$\nu$ and ${\bar \nu}$. Therefore the difference and the sum of
the cross sections are simply summarized as
\begin{equation}
{( {  {d \sigma  } \over {d Q^2  }} )}_{\nu     }^{NC} - {( { {d
\sigma  } \over {d Q^2  }} )}_{ {\bar \nu}    }^{NC} = - { {G_F^2
} \over { 2 \pi }}~ 4 y ( 1 - { 1 \over 2} y ) G_M G_A~,
\end{equation}
\begin{eqnarray}
{( {  {d \sigma  } \over {d Q^2  }} )}_{\nu     }^{NC} + {( { {d
\sigma  } \over {d Q^2  }} )}_{ {\bar \nu}    }^{NC} &=& { {G_F^2
} \over { \pi }} [ { 1 \over 2} y^2 { (G_M )}^2 + ( 1 - y - { { M
}\over {2 E }} y ) { {{ (G_E
)}^2 + { E \over 2 M} y { (G_M )}^2 } \over {1 + {  E \over 2 M} y }} \\
\nonumber & & + ( { 1 \over 2} y^2 + 1 - y + { { M } \over {2 E }}
y ){ (G_A )}^2 ]~.
\end{eqnarray}

Here eq.(9), which is the numerator in the $A_{NC}$, is
proportional to $G_A$. Since the $y$ variable is given as $Q^2 / 2
E_{\nu} M$ in the nucleon rest frame, $y$ is always positive, but
less than $1$ for the energy region, $E_{\nu}< 1 $ GeV and $Q^2 <
1 $ GeV$^2$, considered here. It means that the asymmetry $A_{NC}$
could be very sensitive on the $g_A^s$ value because $A_{NC}$ is
approximated as ${2 y G_M G_A }/{ ( 1 - y) ( G_E^2 + G_A^2) }$ if
$O(y^2)$ and ${  E \over { 2 M}} O(y)$ terms are neglected.
Moreover, the eq.(9) has a positive sign irrespective of the
proton and the neutron. Consequently, $\nu$ cross section is
always larger than that of ${\bar \nu}$ on the nucleon level.

Detailed results on the nucleon level are shown in fig. 1 and 2,
where the results for $g_A^s = -0.19$ and $0.0$ are presented on
the proton (fig.1) and the neutron (fig.2). The cross sections by
the incident $\nu ({\bar \nu})$ on the proton are usually enhanced
in the whole $Q^2$ region by the $g_A^s$, while they are reduced
on the neutron. But the asymmetry on the proton is maximally
decreased in the $Q^2 \sim 0.6 $ GeV$^2$ region about 15 \%, while
on the neutron it is maximally increased in that region. The
ratio, $R_{{\bar \nu}/ \nu}$, on the contrary, shows reversed
behaviors. Therefore, the $g_A^s$ effects can be detected in the
asymmetry $A_{NC}$ around the $Q^2 \sim 0.6 $ GeV$^2$ region, more
clearly than the cross sections.

Since the neutrino energy was not known exactly at the BNL
experiments, one usually defines the flux averaged cross section
\begin{equation}
<\sigma> =  {<{ {d \sigma  } \over {d Q^2  } }>}_{\nu ({\bar
\nu})}^{NC} = {  {\int d E_{\nu ({\bar \nu})}  ({ {d \sigma  } /
{d Q^2  }  })_{\nu ({\bar \nu})}^{NC} \Phi_{\nu ({\bar \nu})}
(E_{\nu ({\bar \nu})} )} \over {{\int d E_{\nu ({\bar \nu})}
\Phi_{\nu ({\bar \nu})} (E_{\nu ({\bar \nu})} )}  }} ~,
\end{equation}
where $\Phi_{\nu ({\bar \nu})} (E_{\nu ({\bar \nu})}) $ is
neutrino and antineutrino energy spectra. The experimental result
at BNL, $R_{{\bar \nu} / {\nu}}^{BNL} = {< {\sigma ( {\bar \nu} p
\rightarrow {\bar \nu} p ) }
> / { < \sigma (\nu p \rightarrow \nu  p )> }}$, turned out to be
about 0.32 \cite{brook,Albe02}. Therefore, the flux averaged
asymmetry $< A_{NC} > = { (<\sigma_{NC}^{\nu}> -
<\sigma_{NC}^{{\bar \nu}}> ) / <(\sigma_{NC}^{\nu}> +
<\sigma_{NC}^{{\bar \nu}} > )}$ is 0.5 on the nucleon level. This
value is approximately consistent with our $A_{NC}$ values in fig.
1 and 2, if they are averaged by $Q^2$.

Now, we investigate whether these asymmetry phenomena still hold
in nuclei or not, and the sensitivity of the cross sections and
the asymmetry in the nucleus on the $g_A^s$ value. For the
application to the  $ \nu - A$ scattering, we briefly summarize
the formula exploited in this paper \cite{kim07}. The four-momenta
of the incident and the outgoing neutrinos (antineutrinos) are
labelled $k_i^{\mu}=(E_i, {\bf k}_i)$, $k_f^{\mu}=(E_f, {\bf
k}_f)$. We choose the nucleus fixed frame where the target nucleus
is seated at the origin of the coordinate system. $p_A^{\mu}=(E_A,
{\bf p}_A)$, $p_{A-1}^{\mu}=(E_{A-1}, {\bf p}_{A-1})$, and
$p^{\mu}=(E_p, {\bf p})$ represent the four-momenta of the target
nucleus, the residual nucleus, and the knocked-out nucleon,
respectively. In the laboratory frame, the inclusive cross section
is given by the contraction between the lepton tensor and the
hadron tensor \cite{udias}
\begin{eqnarray}
{\frac {d\sigma} {dE_f}} = 4\pi^2{\frac {M_N M_{A-1}} {(2\pi)^3
M_A}} \int \sin \theta_l d\theta_l \int \sin \theta_p d\theta_p p
f^{-1}_{rec} \sigma^Z_M [v_L R_L + v_T R_T + h v'_T R'_T ],
\label{cs}
\end{eqnarray}
where $M_N$ is the nucleon mass, $\theta_l$ denotes the scattering
angle of the lepton, and $h=-1$ $(h=+1)$ corresponds to the
helicity of the incident neutrino (antineutrino). The squared
four-momentum transfer is given by $Q^2=q^2 - \omega^2$.
$\sigma^Z_M$ is defined by
\begin{equation}
\sigma^Z_M = \left ( {\frac {G_F \cos (\theta_l/2) E_f M_Z^2}
{{\sqrt 2} \pi (Q^2 + M^2_Z)}} \right ),
\end{equation}
where $G_F$ is the Fermi constant given by $G_F \simeq 1.16639
\times 10^{-11}$ MeV$^{-1}$ and $M_Z$ is the rest mass of
$Z$-boson. The recoil factor $f_{rec}$ is given by
\begin{equation}
f_{rec} = {\frac {E_{A-1}} {M_A}} \left | 1 + {\frac {E_p}
{E_{A-1}}} \left [ 1 - {\frac {{\bf q} \cdot {\bf p}} {p^2}}
\right ] \right |. \end{equation}

For the NC reaction, the coefficients $v$ in eq.(12) are given by
\begin{equation}
v_L=1, \;\;\;\;\;\;\;  v_T=\tan^2 {\frac {\theta_l} {2}} + {\frac
{Q^2} {2q^2}}, \;\;\;\;\;\;\; v'_T=\tan {\frac {\theta_l} {2}}
\left [\tan^2 {\frac {\theta_l} {2}} + {\frac {Q^2} {q^2}} \right
]^{1/2}.
\end{equation}
The corresponding response functions are given by
\begin{eqnarray}
R_L=\left | J^0 - {\frac {\omega} {q}} J^z \right |^2,
\;\;\;\;\;\;\; R_T=|J^x|^2 + |J^y|^2, \;\;\;\;\;\;\; R'_T = 2
{\mbox {Im}}({J^x}J^{y*}).
\end{eqnarray}
Here the $R_T^{'}$, which is a transverse response function for a
target nuclei, just corresponds to the last term in the Eq.(8) on
the nucleon level. We calculate the NC $\nu ({\bar \nu})$
scattering on the target $^{12}$C nucleus in the QE region for two
incident energies, 500 MeV and 1.0 GeV.

Figure \ref{neut-gs} and \ref{anti-gs} exhibit the strange quark
contributions to the cross section by $\nu ({\bar \nu})$
scattering on the $^{12} C$ target nuclei, ${\it i.e.}$ $^{12}$C$(
\nu ( {\bar \nu}   ), \nu^{'} ( {\bar \nu^{'}} ))$. Thick and thin
curves are the results for $g^s_A=-0.19 $ and $ 0.0 $,
respectively. Note that a relativistic optical potential for the
knocked-out nucleon \cite{clark} is taken into account for the
FSI. Detailed discussions about the FSI are presented at our
previous paper \cite{kim07}.

To analyze cross sections for $\nu -^{12} C$ (solid curves) in
fig.\ref{neut-gs}, we present each contribution by the neutron
(dotted curves) and the proton (dashed curves), which includes
knocked-out or excited states of the corresponding nucleon by the
incident $\nu$. The effect of $g^s_A$ for the protons is increased
by 30\% for 500 MeV and 31\% for 1.0 GeV, but for the knocked-out
neutrons it is decreased by 30\% for 500 MeV and 29\% for 1.0 GeV,
maximally. These individual $g_A^s$ effects on each nucleon
resemble exactly those of each nucleon in fig. 1 and 2, {\it
i.e.}, the $g_A^s$ effects by protons increase the cross section
and decreases those of neutrons. These phenomena are quite natural
because the knocked-out nucleons are scattered by the QE processes
and they are main contributions in the inclusive reaction.

However, total net $g_A^s$ effects may severely depend on the
competition between the proton and neutron processes. In the case
of $^{12} C$, the enhancement due to the $g_A^s$ by the proton is
nearly compensated by the neutron process, so that the net effect
of $g_A^s$ increases the cross sections only by about 3\% for 500
MeV and 2\% for 1.0 GeV, maximally. These values are much smaller
than those on the elementary process.

For the incident ${\bar \nu}$ in fig.\ref{anti-gs}, the $g_A^s$
effect enhances the results by 21\% for 500 MeV and by 19\% for
1.0 GeV on the knocked-out protons, but reduces them by 20\% for
500 MeV and 18\% for 1.0 GeV for the neutrons, maximally. The
reduction by the neutron is nearly balanced by the enhancement due
to the proton. Consequently, the net effect of the strange quark
reduces the total cross sections only by 1\% for 500 MeV and 2\%
for 1.0 GeV, maximally, similarly to the $\nu$ case. In specific,
in the $\nu$ case of fig.3, the net effect is nearly
indiscernible.

From these results, in the $\nu - A$ cross sections, the $g^s_A$
effect turned out to contribute more positively to the protons
while it does negatively to the magnitude of the cross sections
for the neutrons. Total net effects in nuclei due to the $g_A^s$
come from the competition of the two knocked-out nucleon processes
in the QE region, whose detailed competition may depend severely
on the target nuclei. In the case of $^{12}$C, the strange quark
effects reduce the cross sections for the ${\nu}$ and the ${\bar
\nu}$, although the resultant effects are only within a few \% by
the cancellation. Therefore, the $g_A^s$ effect in nuclei is too
small to be deduced from the A$( \nu, \nu^{'})$ cross section
itself.

The asymmetry between the $\nu -$ and $ {\bar \nu} - A $
scattering processes is hinged on the third term in Eq.
(\ref{cs}), $R_T^{'}$, due to its helicity dependence. Using the
helicities, the asymmetry between $\nu$ and ${\bar \nu}$ is
written as
\begin{equation}
A_{NC} = {\frac {\sigma(h=-1) - \sigma(h=+1)} {\sigma(h=-1) +
\sigma(h=+1)}}~ , \label{asy}
\end{equation}
where $\sigma$ denotes the differential cross section in Eq.
(\ref{cs}) and $h=-1$ ($h=+1$) represents the helicity of the
incident $\nu ( {\bar \nu})$. Our results for the asymmetry
$A_{NC}$ in the $^{12}$C$( \nu ({\bar \nu}) , \nu' ({\bar
\nu^{'}}))$ reaction are shown in fig.\ref{asy-gs}. For
$E_{\nu}=500 MeV$ case, in the region $T_p = 250 \sim 300$ MeV
region, which just corresponds to $Q^2 ( \sim 2 M T_p) \sim $ 0.6
GeV$^2$ region, the $g_A^s$ effects appear maximally as expected
from the elementary processes. But the amounts of the $g_A^s$
effects, {\it i.e.} the gaps in solid and dashed curves, are much
smaller than the 15 \% on the nucleon level. It is also remarkable
that the $A_{NC}$ value averaged by $T_p$ is as large as that of
the nucleon level, about 0.5 because the $A_{NC}$ averaged by
$T_p$ corresponds to that of $A_{NC}$ averaged by $Q^2$ on the
elementary processes. Therefore $g_A^s$ effects in $A_{NC}$ do not
show any drastic effects in nucleus just like cross sections (the
solid lines in fig. 3 and 4). But, likewise nucleon level,
positive values of $A_{NC}$ indicate that cross sections by $\nu$
is always larger than that of ${\bar \nu}$, at least the QE region
considered here, even in nuclei.

In order to more clearly understand these results, the $g_A^s$
effects on the asymmetry via the proton and the neutron knockout
processes are separately investigated in fig. \ref{asy-gs-pro} and
\ref{asy-gs-neu}. Solid and dashed curves represent the results
with $g^s_A= - 0.19$ and $g^s_A=0.0$, respectively. The $g_A^s$
effects in $A_{NC}$ show a tendency nearly same as those of
nucleons in fig. 1 and 2, {\it i.e.} the proton knockout case is
decreased contrary to the neutron knockout case which is increased
by $g_A^s$ effects. The relatively small effects in fig. 5 turned
out to stem from some cancellation between reduction by protons
and enhancement by neutrons in this inclusive reaction.

If one wants to search for the $g_A^s$ effects in nuclei,
therefore, the asymmetries in the exclusive reactions, $^{12}C (
\nu ({\bar \nu}) , \nu' ({\bar \nu^{'}}) N )$ detecting
knocked-out nucleon, are more efficient tools rather than the
inclusive reactions. In specific, the asymmetry in the $T_p = 200
\sim 250$ MeV region for the knocked-out nucleon in the reaction,
could be more efficient tests on searches of the $g_A^s$ effect.

In conclusion, the $g_A^s$ effect in the NC $\nu ({\bar \nu})$
scattering on the nucleon enhances the cross section for the
proton, but reduces it for the neutron. But in the $\nu - A$
scattering on the QE region both protons' and neutrons'
contributions compensate each other in the cross section, so that
the net effects are sensitive, sometimes nearly indiscernible, on
the nuclear structure because of the possible cancellations. The
effects in the asymmetry between $\nu$ and ${\bar \nu}$ scattering
also show a similar competing mechanism between knocked-out
protons and neutrons. It means that exclusive reactions detecting
knocked-out nucleon, such as $A( \nu ({\bar \nu}) , \nu' ({\bar
\nu^{'}}) N )$, could be more plausible tests for the effect
rather than inclusive reactions, $A( \nu ({\bar \nu}) , \nu'
({\bar \nu^{'}}) )$, because there are no competitions of the
$g_A^s$ effects between the processes via the knockout protons and
the knockout neutrons in the exclusive reaction.

This work was supported by the Soongsil University Research Fund.

\newpage
\begin{figure}
\vskip-3.5cm
\includegraphics[width=0.5\linewidth]{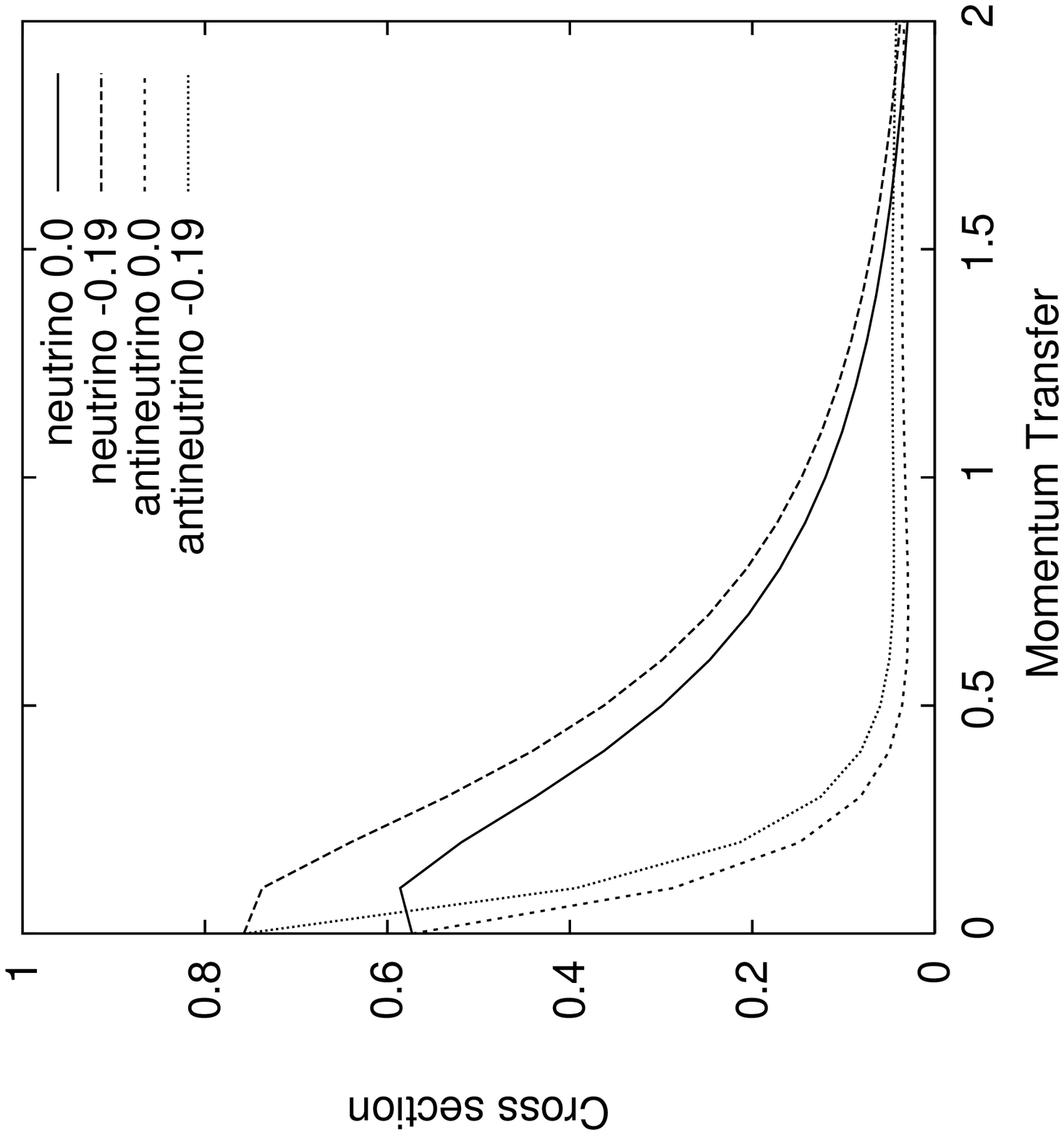}
\includegraphics[width=0.5\linewidth]{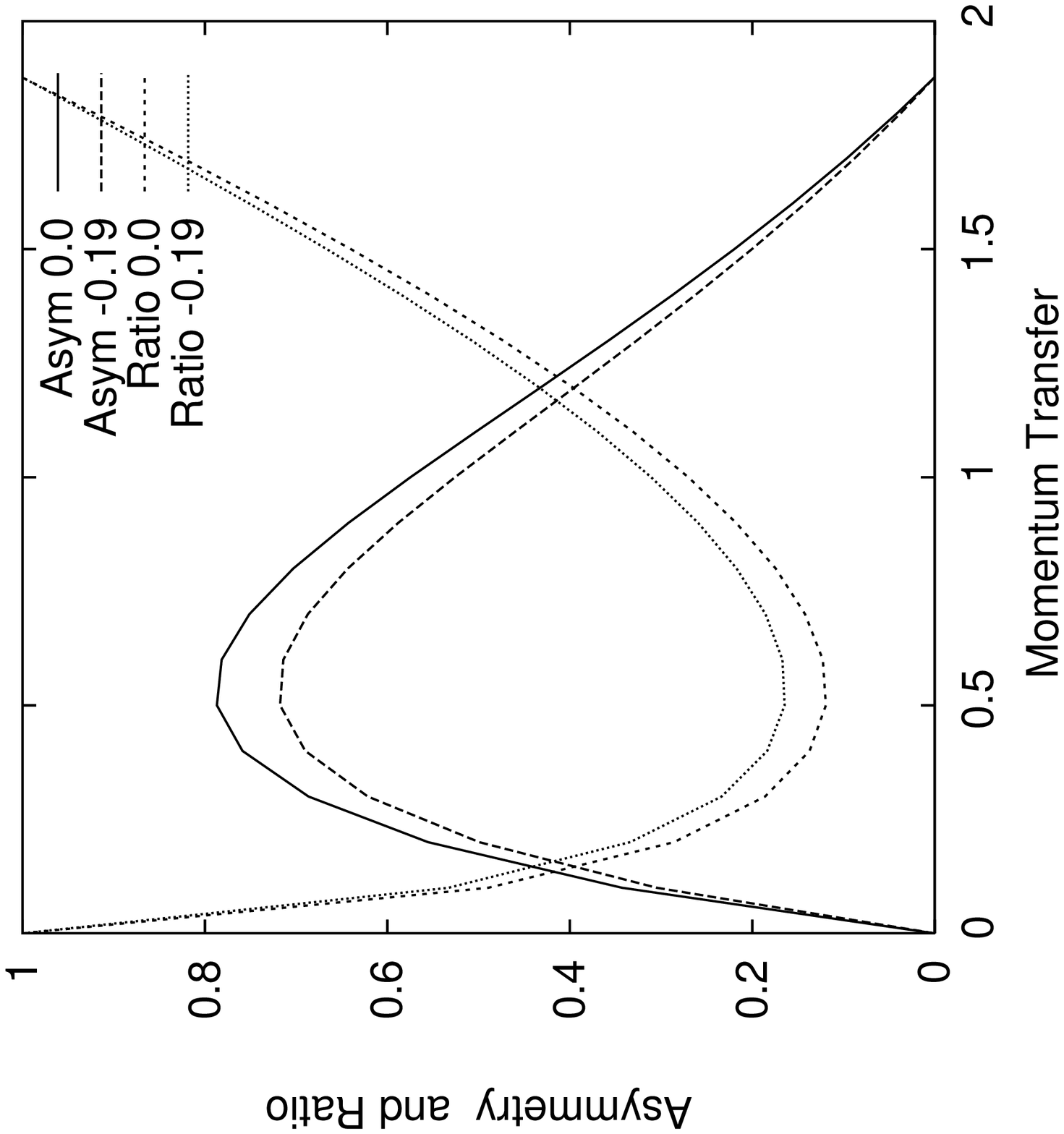}
\caption{$d \sigma / d Q^2$ for $ \nu ({\bar \nu})p \rightarrow
\nu ({\bar \nu}) p $ by the neutral current in an arbitrary unit
(upper part), asymmetry, and ratio $R_{ {\bar \nu } / {\nu} }$
(lower part) on the proton as a function of the incident energy
$E=500$ MeV. They are calculated for $g_A^s = -0.19$ and 0.0
cases, respectively.} \label{neut-gs}
\end{figure}

\newpage
\begin{figure}
\vskip-3.5cm
\includegraphics[width=0.5\linewidth]{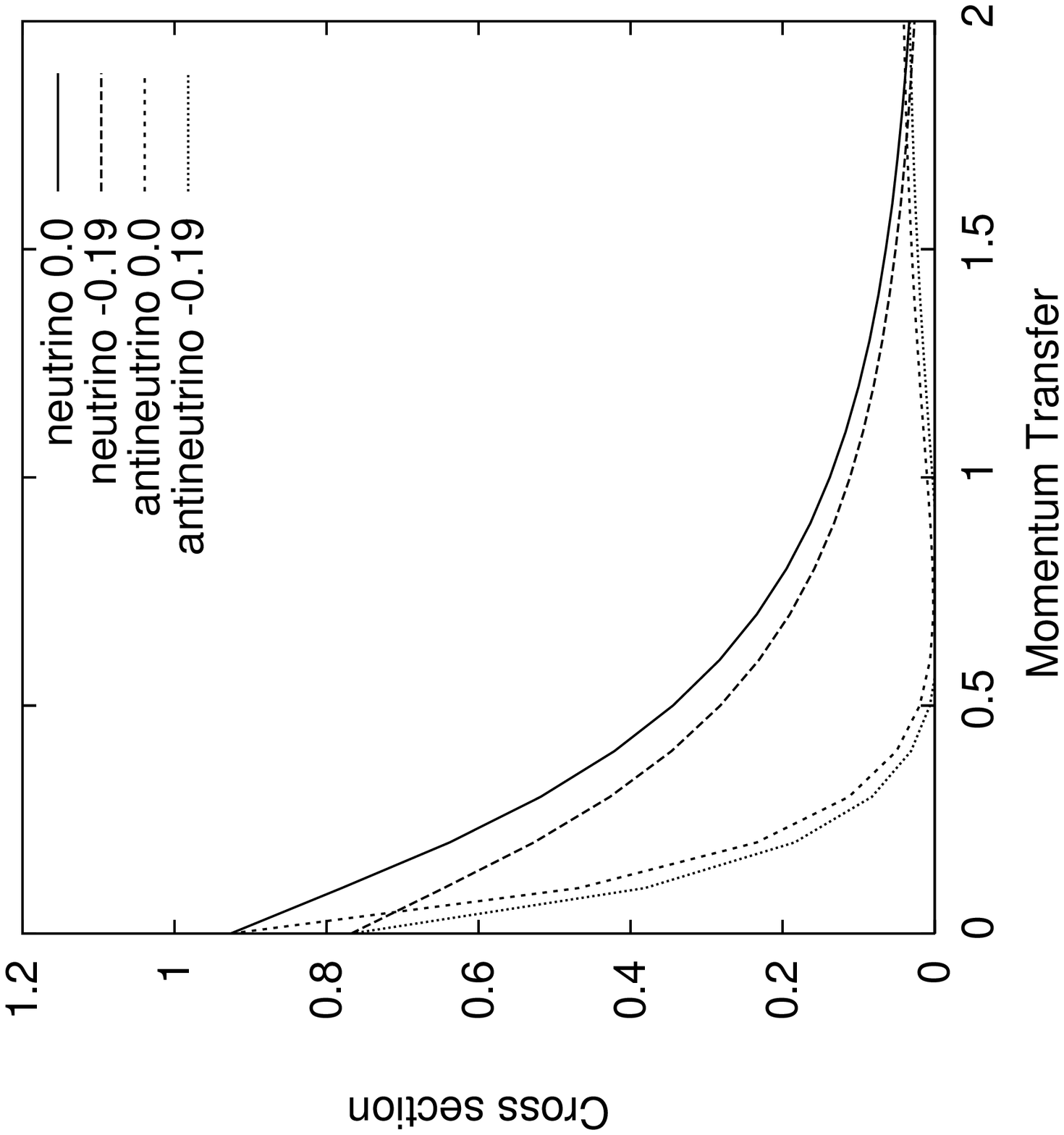}
\includegraphics[width=0.5\linewidth]{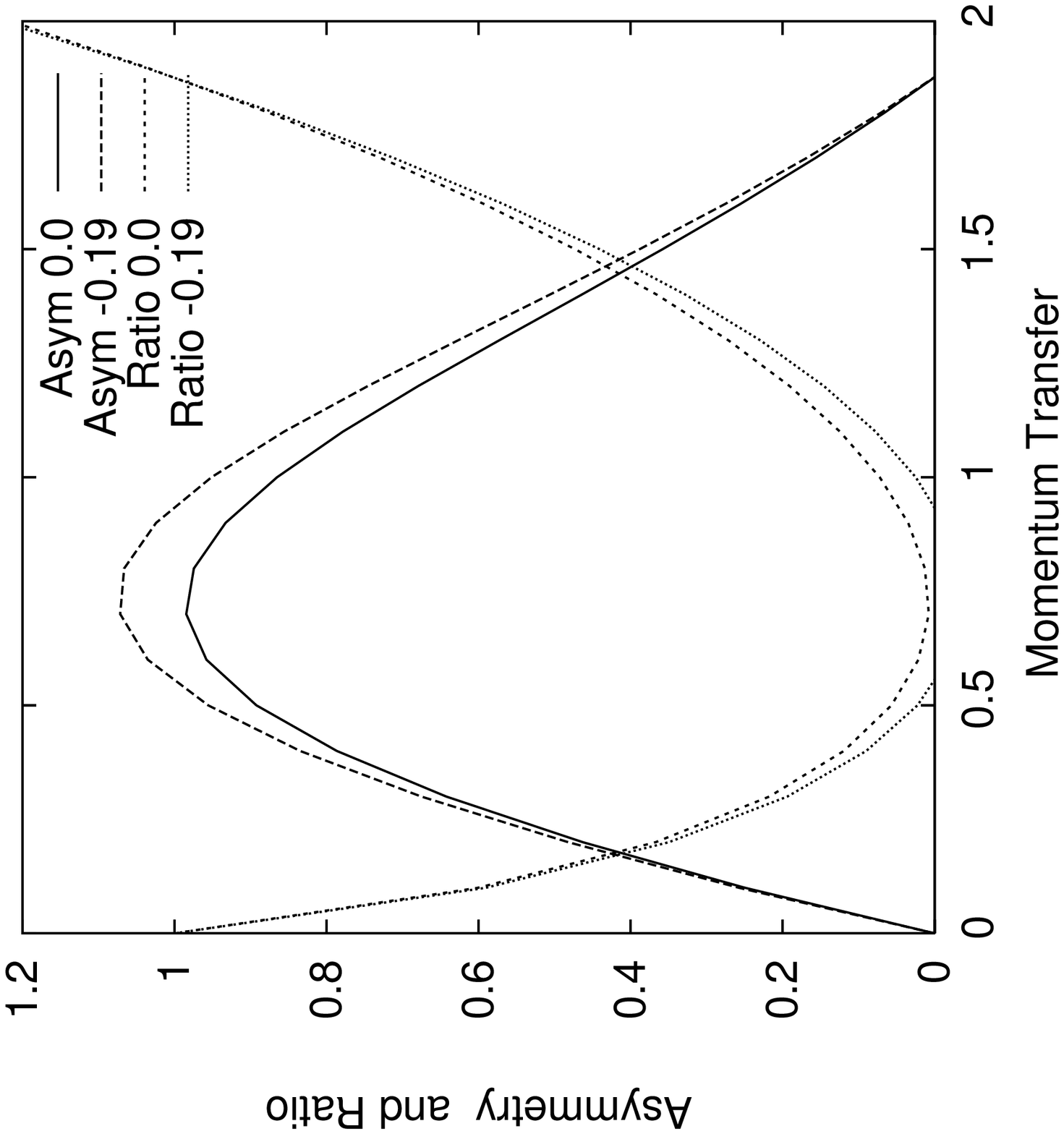}
\caption{The same as figure 1, but for the neutron target}
\label{neut-gs}
\end{figure}

\begin{figure}
\includegraphics[width=0.5\linewidth]{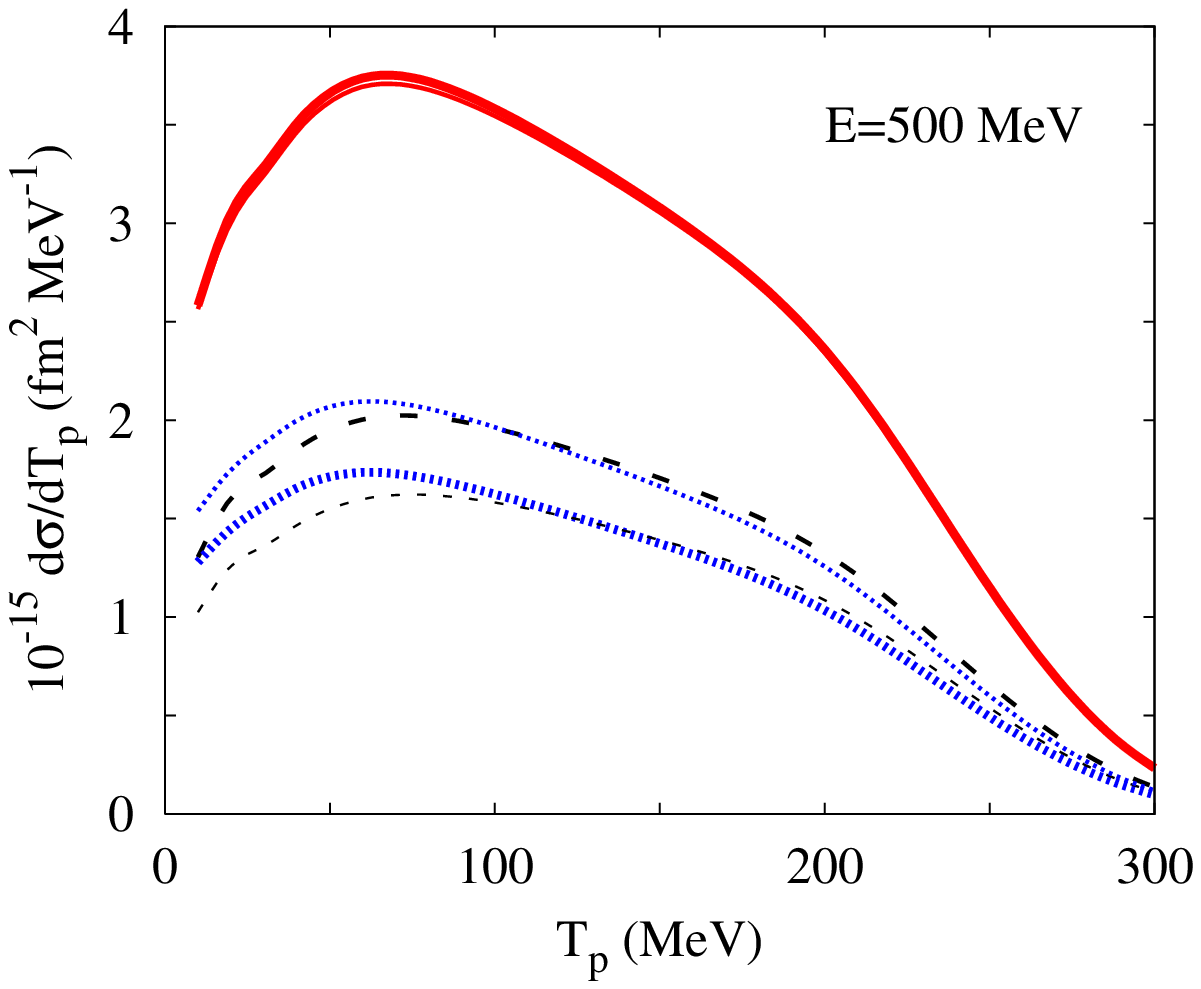}
\includegraphics[width=0.5\linewidth]{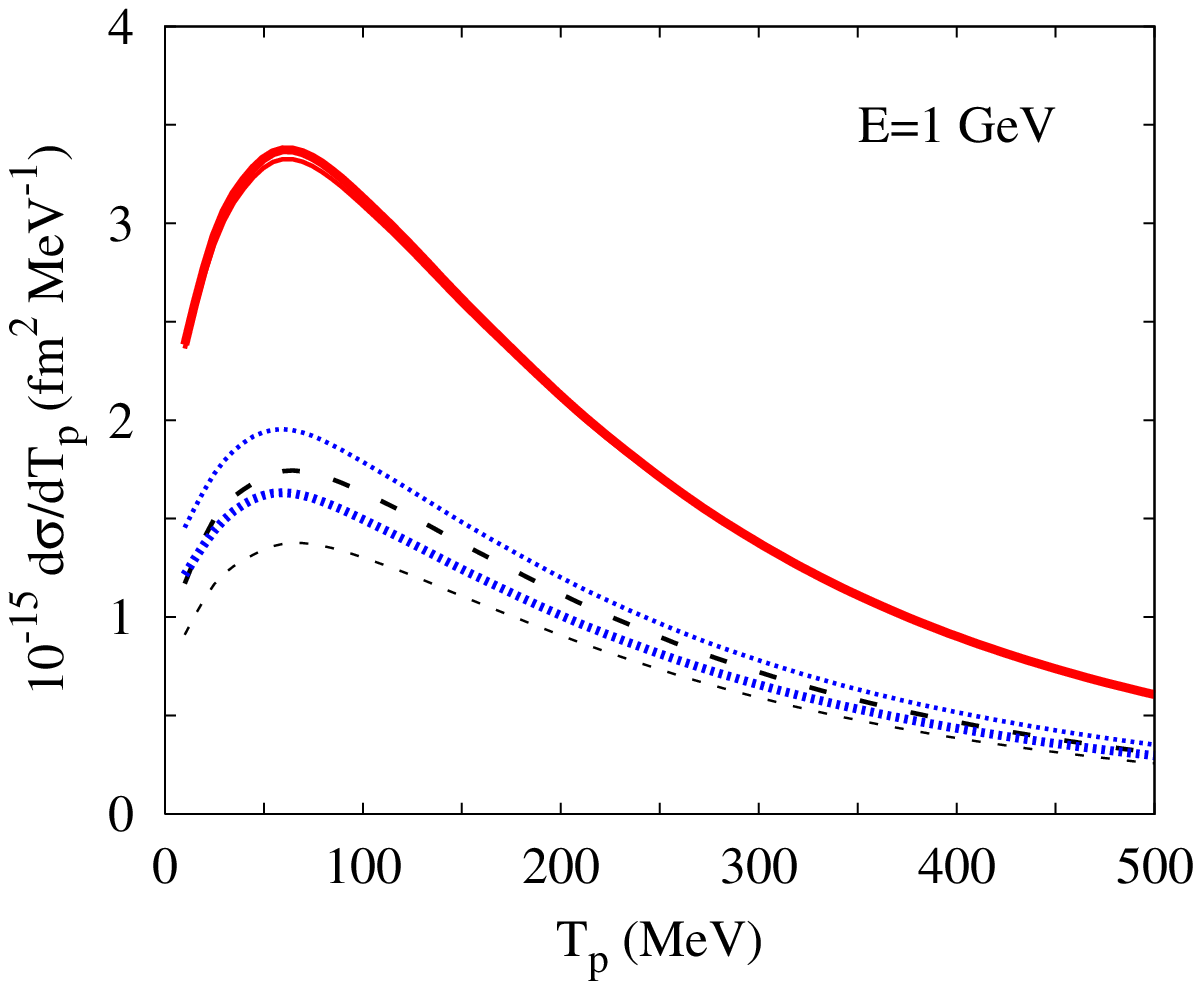}
\caption{Neutral current $^{12}$C($\nu, \nu'$) cross section as a
function of the knocked out nucleon kinetic energy $T_p$ at
incident neutrino energy $E=500$ MeV. Solid curves are the results
for the cross sections, dashed and dotted lines are the
contributions of the proton and the neutron, respectively. Thick
and thin lines are calculations with $g^s_A=-0.19$ and $g^s_A=0$.
The optical potential of the knocked-out nucleon is used for the
final state interaction.} \label{neut-gs}
\end{figure}

\begin{figure}
\includegraphics[width=0.5\linewidth]{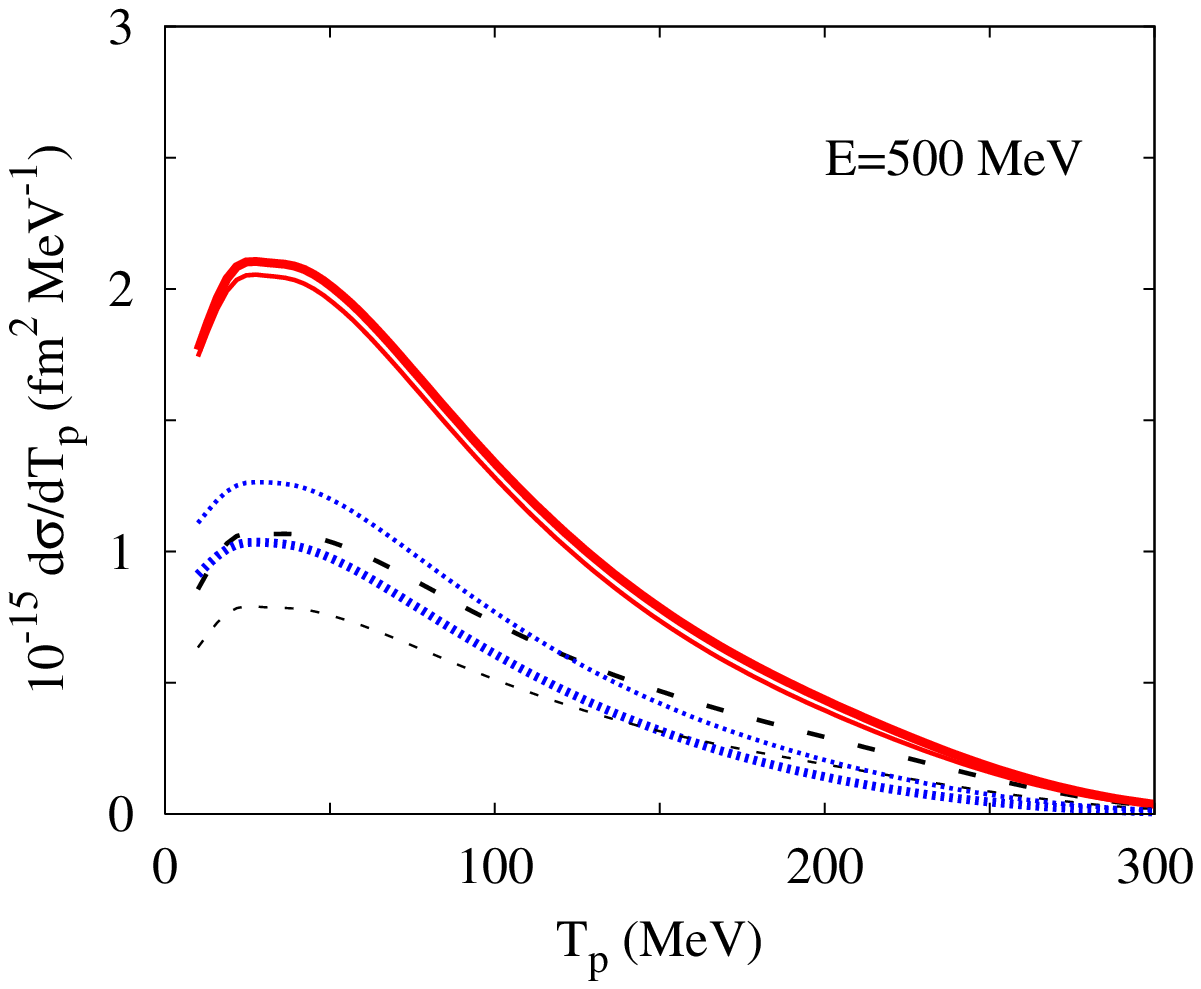}
\includegraphics[width=0.5\linewidth]{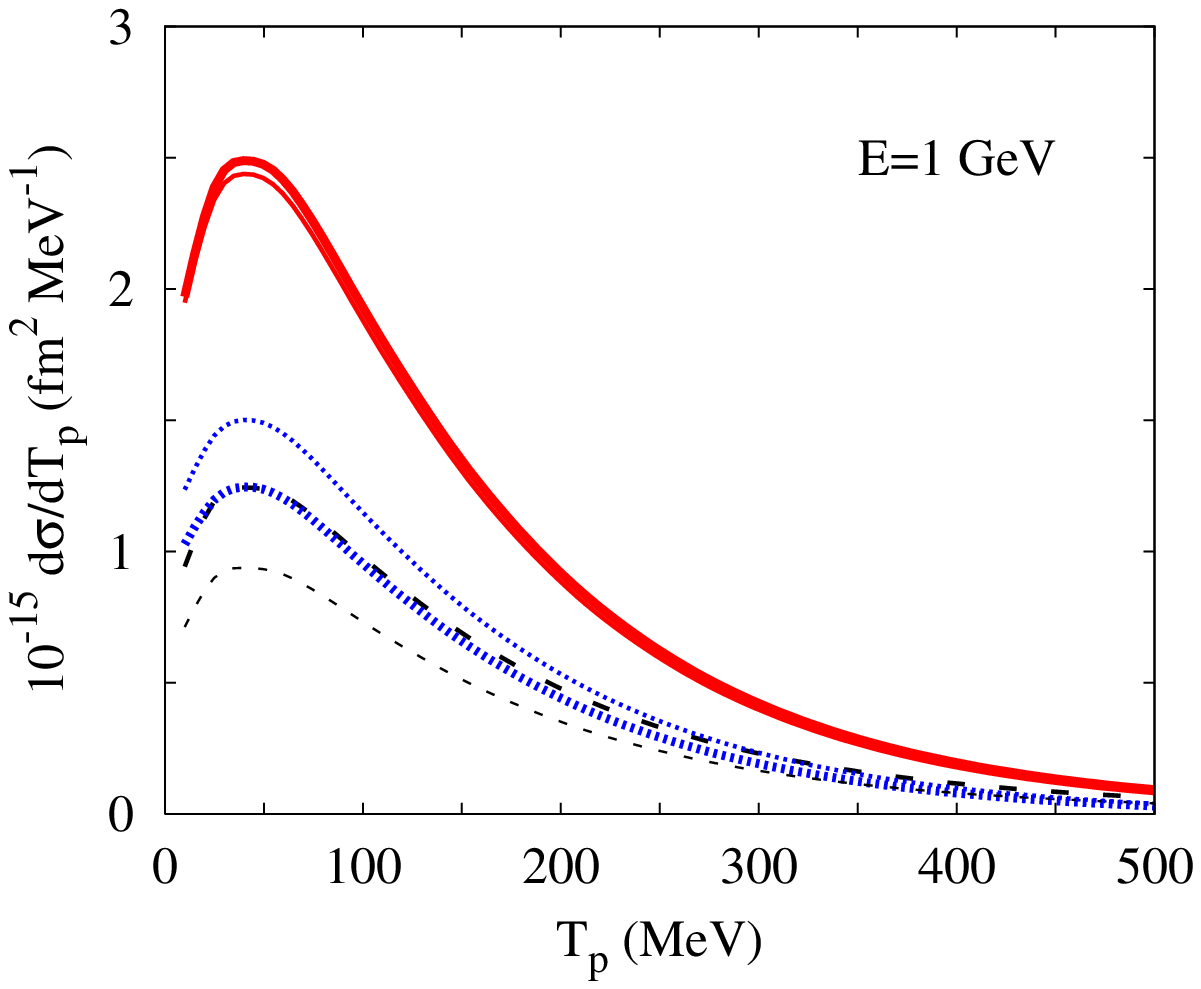}
\caption{The same as Fig. \ref{neut-gs} but for the antineutrino.}
\label{anti-gs}
\end{figure}

\begin{figure}
\includegraphics[width=0.5\linewidth]{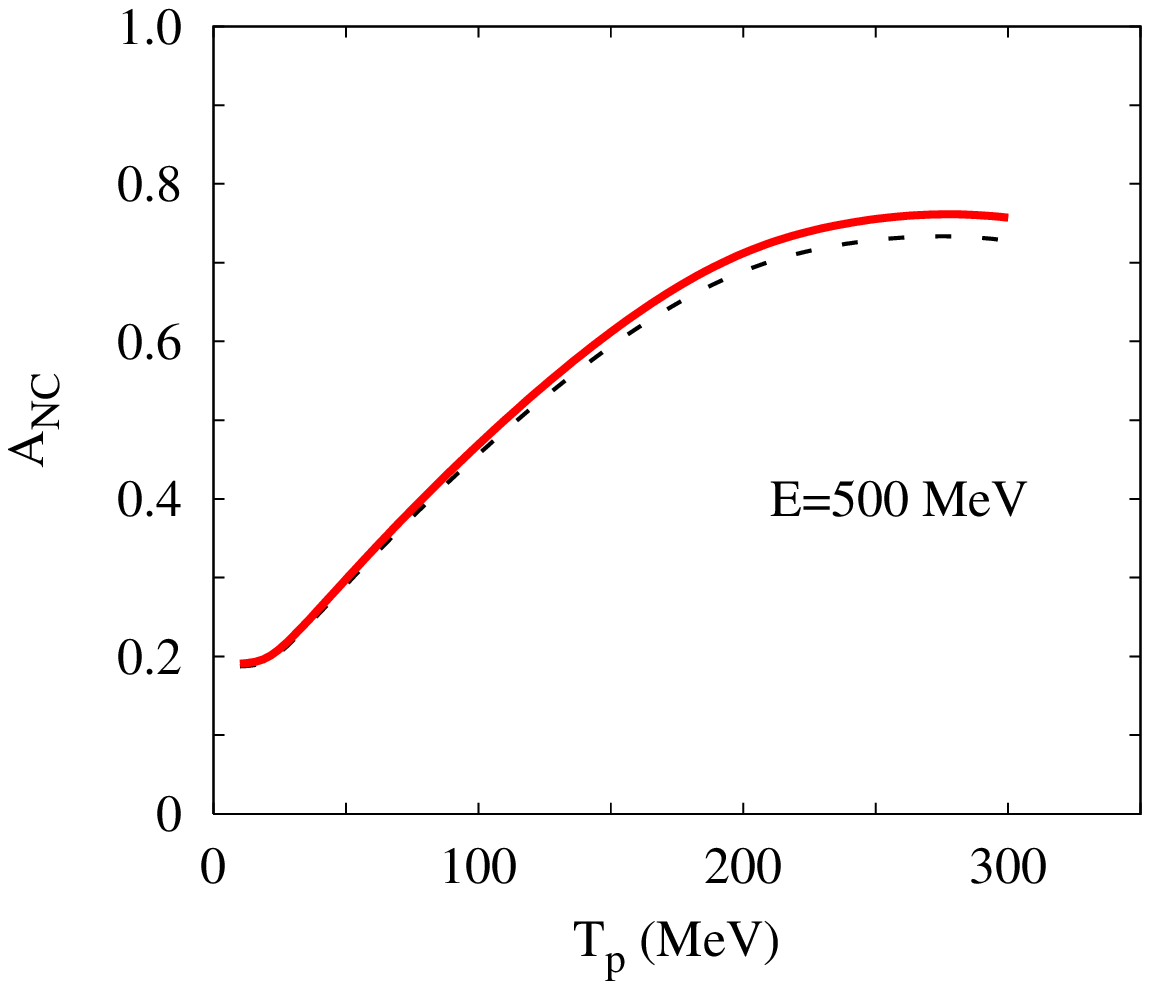}
\includegraphics[width=0.5\linewidth]{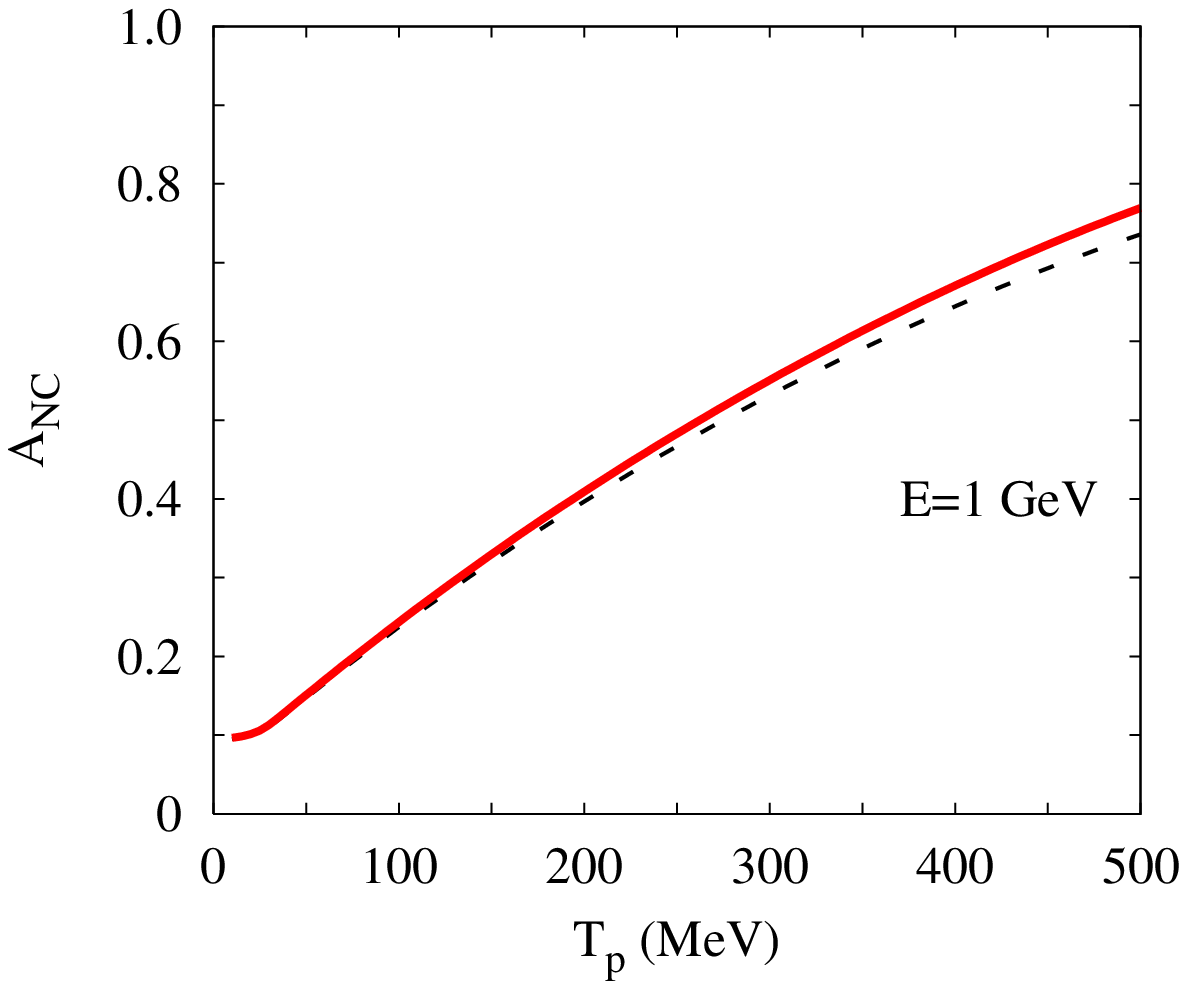}
\caption{The asymmetry as a function of the kinetic energy of the
knocked-out nucleon. Solid and dashed curves represent the results
with $g^s_A=-0.19$ and $g^s_A=0$, respectively.} \label{asy-gs}
\end{figure}

\begin{figure}
\includegraphics[width=0.5\linewidth]{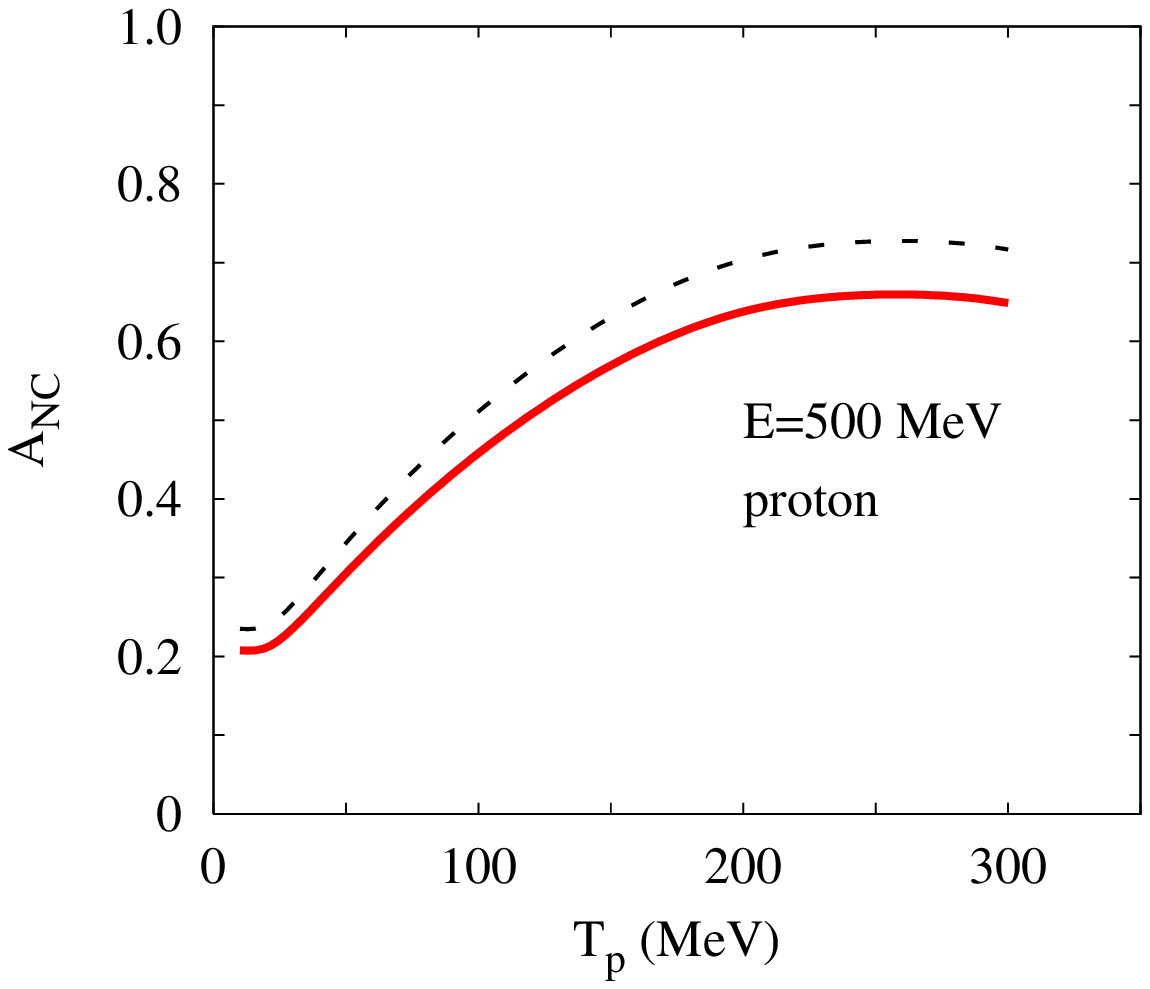}
\includegraphics[width=0.5\linewidth]{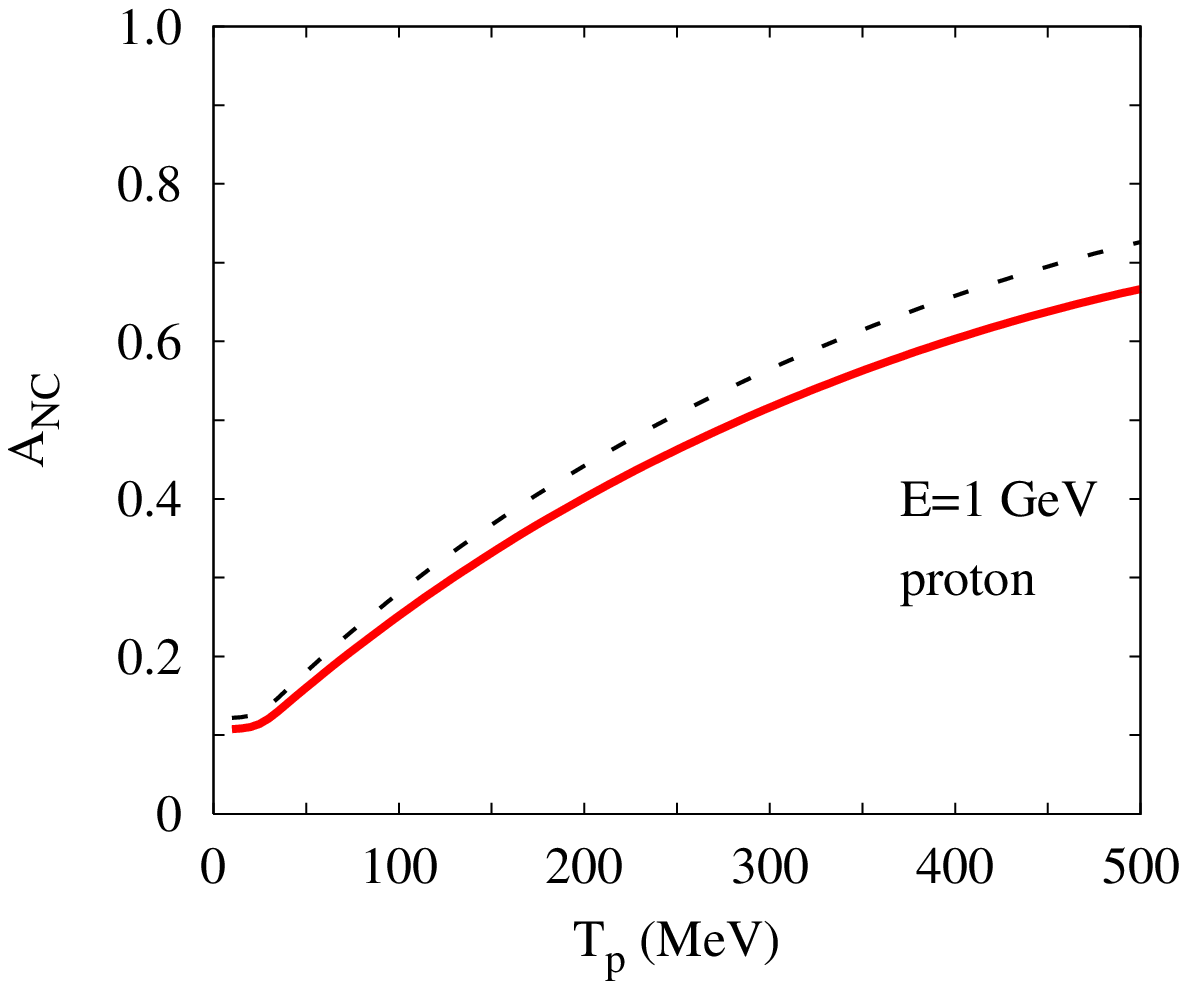}
\caption{The asymmetry as a function of the kinetic energy of the
knocked-out proton with the same kinematics as Fig. \ref{asy-gs}.
Solid and dashed curves represent the results with $g^s_A=-0.19$
and $g^s_A=0$, respectively.} \label{asy-gs-pro}
\end{figure}

\begin{figure}
\includegraphics[width=0.5\linewidth]{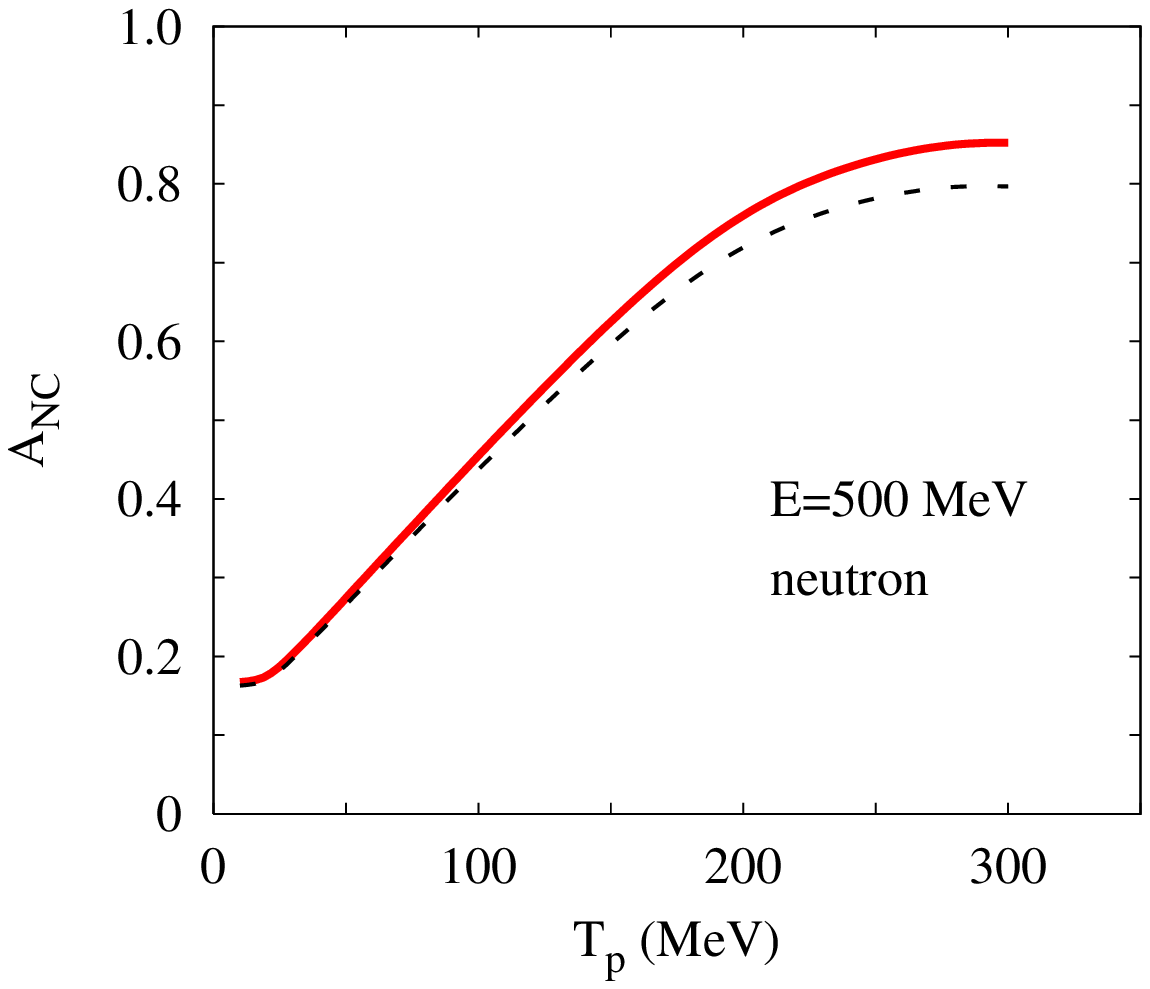}
\includegraphics[width=0.5\linewidth]{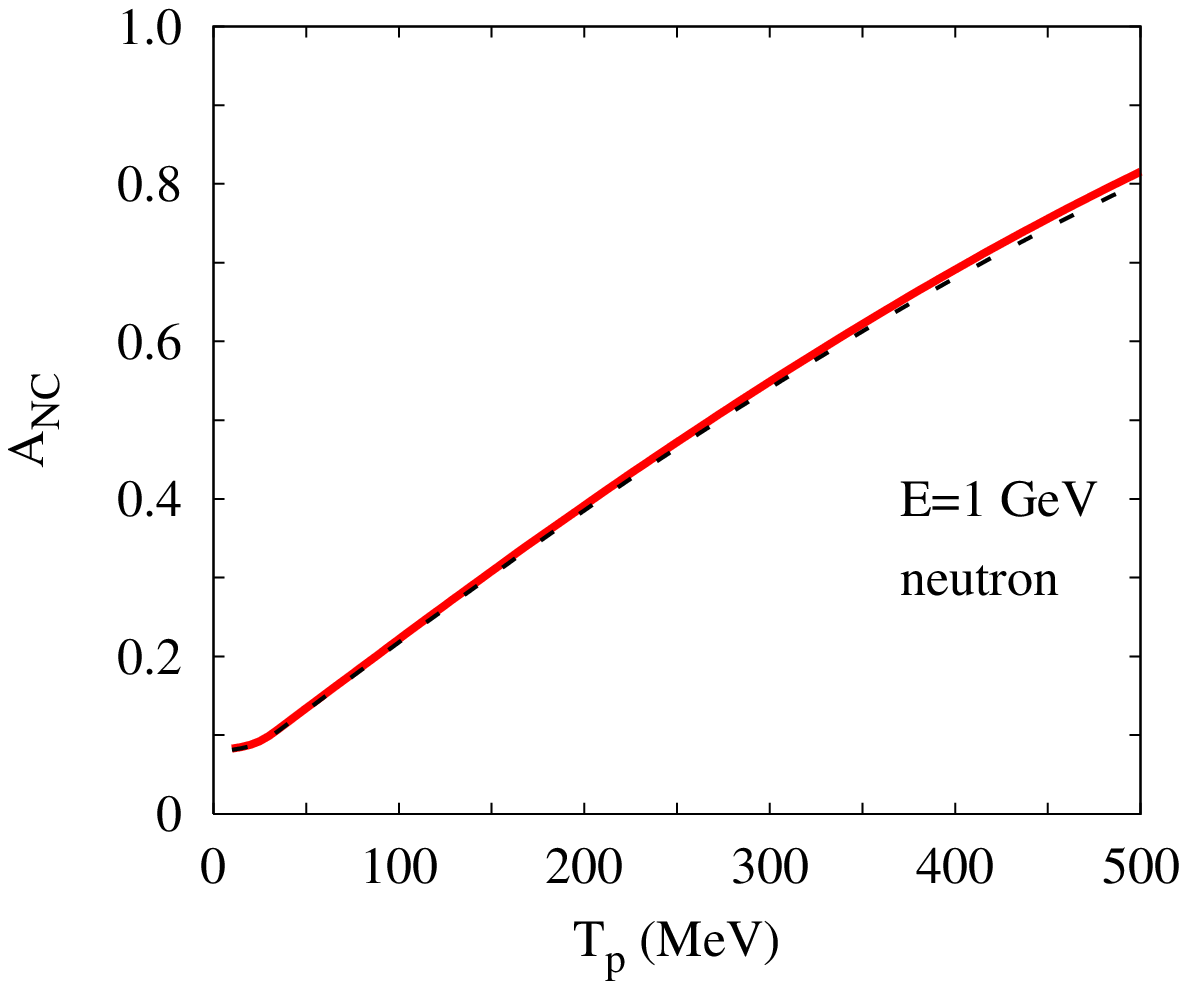}
\caption{The asymmetry as a function of the kinetic energy of the
knocked-out neutron with the same kinematics as Fig. \ref{asy-gs}.
Solid and dashed curves represent the results with $g^s_A=-0.19$
and $g^s_A=0$, respectively. In the case of neutron, the asymmetry
seems to be increased, but the absolute values are decreased
consistently with the neutron case in Fig. 2. The difference in
1.0 GeV case is nearly indiscernible.} \label{asy-gs-neu}
\end{figure}

\end{document}